\documentstyle[aps,preprint]{revtex}

\begin{document}
\tightenlines
\draft
\preprint{FTUV/98-35;~IFIC/98-36}
\title{Nuclear effects in the $\bbox{F_3}$ structure function
}

\author{E.~Marco, E.~Oset and S.K.~Singh\cite{Singhaddress}}

\address{Departamento de F\'{\i}sica Te\'orica and IFIC, \\
Centro Mixto Universidad de Valencia-CSIC, \\
46100 Burjassot (Valencia), Spain}
\date{\today}
\maketitle
\begin{abstract}
By using a relativistic framework and accurate nuclear spectral
functions we evaluate the ratio $F_{3A}/AF_{3N}$ of deep inelastic
neutrino scattering. Parametrizations of this ratio for different
values of $Q^2$ are provided. These results should be useful for taking into
account the nuclear effects in analyses of experimental data in neutrino
reactions in nuclear targets, and test QCD predictions for the nucleon
structure functions. In particular, the size of the nuclear corrections is
of the same order of magnitude as the size of the QCD corrections
to the Gross-Llewellyn Smith sum rule.
\end{abstract}

\pacs{13.15.+g, 24.10.-i, 24.85.+p, 25.30.-c, 25.30.Pt}

\section{Introduction}
The deep inelastic scattering (DIS) experiments done with neutrino beams
can now provide data with an accuracy comparable to the analogous
processes studied with charged leptons \cite{Conrad,SeligmanPRL}.
Due to the low cross sections, for neutrino interaction with matter,
these data are collected mostly from experiments done on heavy nuclei like
iron \cite{SeligmanPRL}.

Since the discovery of the EMC effect \cite{EMC}, the deviation of the
nuclear structure function $F_{2}^l (x)$ with respect to the deuteron,
many theoretical attempts were made to explain it
(for a review see \cite{Bickerstaff,Arneodo,Geesaman}). In the conventional
nuclear physics models, the usual effects considered are Fermi motion
\cite{Bodek} and binding of the nucleons in the nuclei \cite{Akulinichev}
and the contribution of the virtual meson cloud
\cite{Ericson,xlt1}.

For the neutrino processes, the study of possible nuclear effects has
received much less attention, probably because of the lack of accurate
measurements in the past. In present analyses of neutrino experiments,
the high $x$ data ($x>0.7$), where Fermi motion effects are expected to
be important \cite{Bodek}, are generally avoided, especially
in the case of nuclear data \cite{SeligmanPRL}. However, it has been
shown that even in the case of the deuteron, the Fermi motion effects
could be important \cite{Jones}. It is interesting
to note that the older fits of Diemoz et al. \cite{Diemoz},
show that even for lower $x$ ($x\approx 0.5$), the nuclear effects could
be important. Most of the analyses done for the neutrino
data assume the same nuclear corrections as applicable
to the charged lepton experiments in their global fits of the parton
distributions \cite{MRS1,CTEQ}, while allowing for
an overall normalisation of the data. This is not appropriate
for the parity violating structure function $F_{3A}^{\nu} (x)$,
in the neutrino scattering data, which has no counterpart in the charged
lepton case. Even in the case of the structure function $F_{2A}^{\nu} (x)$,
which is related to the $F_{2A}^{l} (x)$, of the EMC effect, the nuclear
corrections would not be the same due to the different contribution
of the sea quarks \cite{Roberts}.

Recently, there have been some calculations of the nuclear effects in
the $F_{3A}^{\nu} (x)$ structure function \cite{Sidorov,Kulagin}.
It is found by Sidorov and Tokarev (ST) \cite{Sidorov} that the nuclear
effects in $F_{3}^{\nu} (x)$ are quite small in the case of the deuteron,
a result similar to the one found in charged lepton scattering
in $F_{2D}^{l} (x)$ \cite{Ciofi,Burov}. Keeping in view the result
that the nuclear effects in $F_{2}^{l} (x)$ are quite different for the
deuteron and heavy nuclei, the assumption that they are the same for
$F_{3}^{\nu} (x)$, as done by ST is therefore not justified. 

In the calculation
of Kulagin \cite{Kulagin}, the iron structure function is evaluated
and sizeable deviations from the deuteron one were found. The effect
on the Gross-Llewellyn Smith (GLS) sum rule \cite{GLS} is on
the contrary very small. This is so because in Kulagin's approach,
the spinors are normalized in order not to affect this sum rule.
It is similar to the prescription introduced by means of the so called flux
factor \cite{Frankfurt} in order to take into account relativistic
effects in non relativistic formalisms. It is interesting to note
that in the fit that the CCFR collaboration made to the $xF_2(x)$
and $xF_3(x)$ neutrino data \cite{SeligmanPRL,SeligmanThesis}, the integral of
the valence quark distributions is not 3,
\cite{SeligmanThesis}. This apparent lack of normalization of the
$F_{3A} (x)$ in the nuclear data is of crucial importance in
order to extract the strong coupling constant, $\alpha_s$, from
the analysis of the GLS sum rule, as done in \cite{Harris}.

Recently, a new calculation of the nuclear effects in the case
of the charged lepton scattering was made \cite{xlt1,xgt1}. In this
calculation a relativistic many-body approach for the nucleon contribution
was developed
in order to avoid the use of the flux factor mentioned before \cite{Frankfurt}.
The meson cloud contribution was found to give relatively important
contributions. Good agreement was found with the experimental data. 
For the structure function $F_{3A} (x)$, the virtual meson cloud gives
no contribution, since it is related to the difference
of the quark and antiquark distributions. In the present paper we
calculate the nuclear effects to $F_{3A} (x)$ in the same formalism
of \cite{xlt1,xgt1}.

The structure of the paper is as follows. We briefly present
the formalism for deep inelastic neutrino scattering in section II.
In section III we show how nuclear effects are calculated.
In section IV we present the results and state
our conclusions.

\section{Deep inelastic neutrino-nucleon scattering}
The invariant matrix element for the Feynman diagram shown in Fig.~1
corresponding to the inelastic neutrino-nucleon scattering is
written as

\begin{equation}	\label{Tinv}
- i T = \left( \frac{i G}{\sqrt{2}}\right) 
\bar{u}_{\mu} (\bbox{k}') 
\gamma^{\alpha} (1 -\gamma_5)u_{\nu} (\bbox{k}) \,
\left(\frac{m_W^2}{q^2-m_W^2}\right)\langle X | J_{\alpha} | N \rangle\,.
\end{equation}
The cross section for the neutrino scattering, $\sigma^{\nu}$, is then given by

\begin{eqnarray}
\sigma^{\nu} =&& \frac{1}{v_{rel}} \frac{2 m_{\nu}}{2 E_{\nu} (\bbox{k})}\,
\frac{2 M}{2 E (\bbox{p})} \, \int \frac{d^3 k'}{(2 \pi)^3} \,
\frac{2 m_{\mu}}{2 E_{\mu} (\bbox{k}')}\nonumber\\
&&\times
\prod^N_{i = 1} \int \frac{d^3 p'_i}{(2 \pi)^3} \, \prod_{l \epsilon f} \,
\left( \frac{2 M'_l}{2 E'_l} \right) \,
\prod_{j \epsilon b}
\left( \frac{1}{2 \omega'_j} \right)
\bar{\sum} \sum |T|^2 \nonumber\\
&&\times	\label{cross}
(2 \pi)^4 \delta^4 (p + k - k' - \sum^N_{i = 1} p'_i)\,,
\end{eqnarray}

\noindent
where $f$ stands for
fermions and $b$ for bosons in
the final state $X$. The index $i$ is split
in $l, j$ for fermions and bosons respectively.

By inserting expression (\ref{Tinv}) in Eq.~(\ref{cross}), summing over
the final spins and averaging over the initial spins,
the cross section can be expressed in the rest frame of the nucleon,
with $\Omega', E'$ referring to the outgoing muon, as

\begin{equation} 	\label{dif_cross}
\frac{d^2 \sigma^{\nu}}{d \Omega' d E'} 
= \frac{G^2}{(2\pi)^2} \; \frac{|\bbox{k}'|}{|\bbox{k}|} \;
\left(\frac{m_W^2}{q^2-m_W^2}\right)^2
[L^{\alpha \beta} + L_5^{\alpha \beta}]
\; W_{\alpha \beta}^{\nu}\,,
\end{equation}
where the lepton tensors, $L^{\alpha \beta}$ and $L_5^{\alpha \beta}$
are given by

$$
L^{\alpha \beta}=k^{\alpha}k'^{\beta}+k^{\beta}k'^{\alpha}
-(kk')g^{\alpha \beta}\,,
$$

\begin{equation}
L_5^{\alpha \beta}=-i \epsilon^{\alpha \beta \rho \sigma} k_{\rho} 
k'_{\sigma}\,,
\end{equation}
and the hadronic tensor is defined as

\begin{eqnarray}
W^{\nu}_{\alpha \beta} =&& \frac{1}{2\pi} \bar{\sum_{s_p}} \;
\sum_X \; \sum_{s_i} \prod^N_{i = 1}
\; \int \frac{d^3 p'_i}{(2 \pi)^3} \; \prod_{l \epsilon f} \;
\left(
\frac{2 M'_l}{2 E'_l}
\right) \;
\prod_{j \epsilon b} \;
\left(
\frac{1}{2 \omega'_j}
\right)\nonumber\\
&&\times
\langle X | J_{\alpha} | N \rangle 
\langle X | J_{\beta} | N \rangle^* (2 \pi)^4 \delta^4 (p + q - 
\sum^N_{i = 1} p'_i)\,,
\end{eqnarray}

\noindent
where $q$ is the momentum of the virtual $W$, $s_p$ the spin of the
nucleon and $s_i$ the spin of the fermions in $X$. 

In the case of antineutrino scattering, the invariant $T$ matrix is given
by

\begin{equation}	\label{Tinv_anti}
- i T = \left( \frac{i G}{\sqrt{2}} \right) \bar{v}_{\nu} (\bbox{k}) 
\gamma^{\alpha} (1 -\gamma_5) v_{\mu} (\bbox{k'}) \,
\left(\frac{m_W^2}{q^2-m_W^2}\right)
\langle X | J^{\dagger}_{\alpha} | N \rangle
\end{equation}
and the cross section reads

\begin{equation}	\label{dif_cross2}
\frac{d^2 \sigma^{\bar{\nu}}}{d \Omega' d E'} 
= \frac{G^2}{(2\pi)^2} \; \frac{|\bbox{k}'|}{|\bbox{k}|} \;
\left(\frac{m_W^2}{q^2-m_W^2}\right)^2
[L^{\alpha \beta} - L_5^{\alpha \beta}]
\; W_{\alpha \beta}^{\bar{\nu}}\,.
\end{equation}
The usual convention is to express the hadronic tensor as \cite{Bilenky}

\begin{eqnarray}
W^{\nu (\bar{\nu})}_{\alpha \beta} =&& 
\left( \frac{q_{\alpha} q_{\beta}}{q^2} - g_{\alpha \beta} \right) \;
W_1^{\nu (\bar{\nu})}
+ \frac{1}{M^2}\left( p_{\alpha} - \frac{p q}{q^2} \; q_{\alpha} \right)
\left( p_{\beta} - \frac{p . q}{q^2} \; q_{\beta} \right)
W_2^{\nu (\bar{\nu})}\nonumber\\
&&-\frac{i}{2M^2} \epsilon_{\alpha \beta \rho \sigma} p^{\rho} q^{\sigma}
W_3^{\nu (\bar{\nu})} + \frac{1}{M^2} q_{\alpha} q_{\beta}
W_4^{\nu (\bar{\nu})}\nonumber\\
&& \label{had_ten}
+\frac{1}{M^2} (p_{\alpha} q_{\beta} + q_{\alpha} p_{\beta})
W_5^{\nu (\bar{\nu})}
+\frac{i}{M^2} (p_{\alpha} q_{\beta} - q_{\alpha} p_{\beta})
W_6^{\nu (\bar{\nu})}\,,
\end{eqnarray}
where $W_i^{\nu (\bar{\nu})}$ are the structure functions, which depend
on the scalars $q^2$ and $pq$. The terms depending on $W_4$, $W_5$
and $W_6$ in Eq.~(\ref{had_ten}) do not contribute to the cross
section in the DIS \cite{Bilenky}.

Defining the variables

\begin{equation}	\label{Bj_var}
Q^2=-q^2; \quad \nu=\frac{pq}{M}; \quad x=\frac{Q^2}{2M\nu};
\quad y=\frac{\nu}{E_{\nu}(\bbox{k})}\,,
\end{equation}
we can write in the Bjorken limit

\begin{equation}
\frac{d^2 \sigma^{\nu(\bar{\nu})}}{d x d y} =
\frac{G^2ME_{\nu}(\bbox{k}) }{\pi}
\left\{xy^2 F_1^{\nu(\bar{\nu})} (x) 
+ \left(1-y-\frac{xyM}{2 E_{\nu}(\bbox{k})}\right) F_2^{\nu(\bar{\nu})} (x)
\pm  xy(1-y/2)F_3^{\nu(\bar{\nu})} (x)
\right\}\,,
\end{equation}
where the $+$ ($-$) sign stands for the neutrino (antineutrino)
cross section, and the $F_i^{\nu(\bar{\nu})} (x)$ are dimensionless
structure functions defined as 

\begin{equation}
F_1^{\nu(\bar{\nu})} = M W_1^{\nu(\bar{\nu})} ;\quad
F_2^{\nu(\bar{\nu})} = \nu W_2^{\nu(\bar{\nu})} ;\quad
F_3^{\nu(\bar{\nu})} = \nu W_3^{\nu(\bar{\nu})} \,.
\end{equation}
In the quark parton model, the structure functions are expressed in terms
of the quark distributions, $q(x)$ and $\bar{q}(x)$.
Using Callan-Gross relation, these structure functions are given as

$$
2xF^{\nu}_{1N}=2xF^{\bar{\nu}}_{1N}=F^{\nu}_{2N}=F^{\bar{\nu}}_{2N}
=x(q(x)+\bar{q}(x))\,,
$$

$$
F^{\nu}_{3N}= q(x) - \bar{q}(x) + 2s(x) -2 c(x)\,,
$$

$$
F^{\bar{\nu}}_{3N} = q(x) - \bar{q}(x) - 2s(x) +2 c(x)\,,
$$

\begin{equation}	\label{struc}
F_3(x)=\frac{1}{2}[F^{\nu}_{3N} (x) + F^{\bar{\nu}}_{3N}(x)]=q(x) -\bar{q}(x)=
u_v(x)+d_v(x)\,.
\end{equation}
Here 

\begin{eqnarray}
q(x)&=&u(x)+d(x)+s(x)+c(x)\nonumber\\
\bar{q}(x)&=&\bar{u}(x)+\bar{d}(x)+\bar{s}(x)+\bar{c}(x)\,.
\end{eqnarray}
The neutrino structure function $F_{2N}^{\nu}$ is related with
the analogous one for charged lepton scattering $F_{2N}^{l}$ by

\begin{equation}
\frac{F_{2N}^{l}}{F_{2N}^{\nu}} = \frac{5}{18} 
\left(1 - \frac{3}{5} \frac{s(x)+\bar{s}(x)-c(x)-\bar{c}(x)}{q(x)+\bar{q}(x)}
\right)\,.
\end{equation}
In the deep inelastic region, quark distributions
satisfy sum rules in order to give the correct charge,
strangeness and charm of the proton and the neutron. This implies the
constraint

\begin{equation}
\int_0^{1} F_3(x) \,dx =3\,,
\end{equation}
known as the Gross-Llewellyn Smith (GSL) sum rule \cite{GLS}.

QCD corrections modify the expression of the structure functions in
terms of the quark distributions, Eqs.~(\ref{struc}). At leading order,
Eqs.~(\ref{struc}) retain the same structure 
but the quark distributions evolve with $Q^2$
according to the DGLAP (Dokshitzer-Gribov-Lipatov-Altarelli-Parisi)
evolution equations \cite{DGLAP}. When higher orders are considered,
Eqs.~(\ref{struc}) are modified including explicitly the strong
coupling constant, $\alpha_S(Q^2)$.

The GLS sum rule is modified and in the NLO
approximation is given by

\begin{equation}
\int_0^{1} F_3(x) \,dx =3(1-\frac{\alpha_S(Q^2)}{\pi})\,.
\end{equation}
Further higher order corrections up to order $\alpha^3_S(Q^2)$ can be found
in \cite{Larin}.

\section{Nuclear effects in neutrino scattering}
In order to calculate the neutrino-nucleus cross section
we first evaluate the related neutrino self-energy in the medium,
Fig.~2. The self-energy is given by

\begin{eqnarray}
- i \Sigma (k) =&& -\frac{G}{\sqrt{2}}
\int \frac{d^4 q}{(2 \pi)^4} \; 
\bar{u}_{\nu} (\bbox{k})  \gamma_{\beta} (1-\gamma_5)
\; i \frac{\not \! k' + m_{\mu}}{k'^{2} - m^2_{\mu} + i \epsilon} \; 
\gamma_{\alpha} (1-\gamma_5) u_{\nu} (\bbox{k})\nonumber\\
&&\times	\label{Neu_self1}
\left(\frac{-im_W }{q^2-m_W^2}\right)^2 \; (- i) \; \Pi^{\alpha \beta} (q)\,,
\end{eqnarray}
where $\Pi^{\alpha \beta} (q)$ is the $W$ self-energy in the medium.

Eq.~(\ref{Neu_self1}) can be rewritten in the form

\begin{equation}	\label{Neu_self2}
\Sigma(k)= \frac{i G}{\sqrt{2}} \frac{4}{m_{\nu}}
\int \frac{d^4 q}{(2 \pi)^4} \; 
\frac{L^{\alpha \beta} + L_5^{\alpha \beta}}
{k'^{2} - m^2_{\mu} + i \epsilon}
\left(\frac{m_W}{q^2-m_W^2}\right)^2 \; \Pi_{\alpha \beta} (q)\,.
\end{equation}
The probability per unit time for the neutrino to collide with nucleons
when travelling through nuclear matter is \cite{xlt1}

\begin{equation}
\Gamma (k) = - \frac{2 m_{\nu}}{E_{\nu} (\bbox{k})} \; 
\mbox{Im} \; \Sigma (k)\,,
\end{equation}
and the cross section for an element of volume $d^3 r$ in the nucleus is

\begin{eqnarray}
d \sigma & = & \Gamma d t d S = \Gamma \frac{dt}{dl} \;
dl dS = \frac{\Gamma}{v} \; d^3 r =\nonumber\\
& = & \Gamma \; \frac{E_{\nu} (\bbox{k})}{|\bbox{k}|}
 \; d^3 r = - \frac{2 m_{\nu}}{|\bbox{k}|} \;
\hbox{Im} \; \Sigma \; d^3 r \,.
\end{eqnarray}
$\hbox{Im} \; \Sigma$ can be easily evaluated from Eq.~(\ref{Neu_self2})
by means of the Cutkosky rules \cite{Itzykson}

\begin{equation}	\label{Cutkosky}
\begin{array}{lll}
\Sigma (k) & \rightarrow & 2 i \; \hbox{Im} \; \Sigma (k) \\
D (k') & \rightarrow & 2 i \theta (k'^0) \;\hbox{Im} \; D (k') \; 
\hbox{(boson propagator)} \\
\Pi^{\mu \nu} (q) & \rightarrow & 2 i \theta (q^0) \;\hbox{Im} \;
 \Pi^{\mu \nu} (q) \\
G (p) & \rightarrow & 2 i \theta (p^0) \;\hbox{Im} \; G (p) \; 
\hbox{(fermion propagator)} 
\end{array}
\end{equation}
and we get

\begin{equation}	\label{cross_nuclear}
\frac{d^2 \sigma^{\nu}}{d \Omega' d E'} 
= - \frac{G}{\sqrt{2}}\frac{4}{(2\pi)^3} \frac{|\bbox{k}'|}{|\bbox{k}|}
\left(\frac{m_W}{q^2-m_W^2}\right)^2
[L^{\alpha \beta} + L_5^{\alpha \beta}]
\; \int d^3 r \; \hbox{Im} \; \Pi_{\alpha \beta} (q) \,.
\end{equation}
Comparing Eq.~(\ref{dif_cross}) and Eq.~(\ref{cross_nuclear})
we see that

\begin{equation}	\label{W_A}
W^{\alpha \beta}_A (q) = - \frac{\sqrt{2}}{\pi} 
\; \frac{1}{G m_W^2}  \; \int d^3 r \; 
\hbox{Im} \; \Pi^{\alpha \beta} (q)\,.
\end{equation}
Next, we evaluate the $W$ self-energy in the medium

\begin{eqnarray}
- i \Pi^{\alpha \beta} (q) =&&
( - ) \; \int \frac{d^4 p}{(2 \pi)^4} \; i G (p) \;
\sum_X \; \sum_{s_p, s_i} \prod^N_{i = 1}
\int \frac{d^4 p'_i}{(2 \pi)^4}\nonumber\\
&&\times
\prod_l i G_l (p'_l) \prod_j \; i D_j (p'_j) 
\left( \frac{-G m_W^2}{\sqrt{2}} \right)
\langle X | J^{\alpha} | N \rangle \langle X | J^{\beta} | N \rangle^*
\nonumber\\
&&\times \label{selfW}
(2 \pi)^4 \delta^4 (q + p - \Sigma^N_{i = 1} p'_i)\,,
\end{eqnarray}
where we have a minus sign because of the necessary fermion loop.
In the antineutrino case the expressions obtained are very similar.
$L_5^{\alpha \beta}$ appears, as in Eq.~(\ref{dif_cross2}), with
a minus sign in front and in the $W$ self-energy, Eq.~(\ref{selfW}),
we have $\langle X | J^{\dagger}_{\alpha} | N \rangle$, instead
of $\langle X | J_{\alpha} | N \rangle$.
From now on we will always speak of the average of neutrino and
antineutrino structure functions and will omit the superscripts
$\nu$ and $\bar{\nu}$.
For the nucleon propagator in the medium, $G(p)$, we take a
relativistic version \cite{xlt1}, which can be written as

\begin{eqnarray}
G (p^0, \bbox{p}) =&& \frac{M}{E (\bbox{p})} 
\sum_r u_r (\bbox{p}) \bar{u}_r (\bbox{p})
\left[\int^{\mu}_{- \infty} d \, \omega 
\frac{S_h (\omega, \bbox{p})}{p^0 - \omega - i \eta}
\right.\nonumber\\
&&	\label{medium_prop}
\left.
+ \int^{\infty}_{\mu} d \, \omega 
\frac{S_p (\omega, \bbox{p})}{p^0 - \omega + i \eta}\right]\,,
\end{eqnarray}
$S_h (\omega, \bbox{p})$ and $S_p (\omega, \bbox{p})$ being the hole
and particle spectral functions respectively, which are taken from
the work of \cite{xlt1,Fernandez}.

We use the local density approximation in which the spectral functions
depend on the density of the point of the nucleus at which they are
evaluated. In our formalism we use spectral functions for symmetric
nuclear matter. The normalization of the hole spectral function
is given by

\begin{equation}
4 \int d^3 r \;  \int \frac{d^3 p}{(2 \pi)^3} 
\int^{\mu}_{- \infty} \; S_h (\omega, \bbox{p}, k_F (\bbox{r})) 
\; d \omega = A\,,
\end{equation}
where $k_F (\bbox{r}) = [3 \pi^2 \rho (\bbox{r}) / 2]^{1/3}$ is the local Fermi
momentum at the point $\bbox{r}$. All our calculations are done for $^{56}$Fe.
The density for this nucleus is expressed as a two Fermi parameter distribution
given in \cite{Vries}.

We now calculate $\hbox{Im} \; \Pi^{\alpha \beta} (q)$ by using
Cutkosky rules, Eq.~(\ref{Cutkosky}), the expression of
Eq.~(\ref{medium_prop}) for the nucleon propagator in the medium, free
propagators for particles in the final state and by means of
Eq.~(\ref{W_A}) we have the hadronic tensor

\begin{equation}	\label{conv_WA}
W^{\alpha \beta}_{A} = 4 \int \, d^3 r \, \int \frac{d^3 p}{(2 \pi)^3} \, 
\frac{M}{E (\bbox{p})} \, \int^{\mu}_{- \infty} d p^0 S_h (p^0, \bbox{p})
W^{\alpha \beta}_{N} (p, q)\,.
\end{equation}

In order to evaluate $F_{3A}$, we calculate the components $xy$
on both sides of Eq.~(\ref{conv_WA}). We have, using $\epsilon_{0123}=1$,
and taking $\bbox{q}$ along the $z$ axis, as usual,
\begin{equation}
W^{xy}_{A}= -\frac{i}{2M_A} q_z W_{3A}\,,
\end{equation}
and for the right hand side we will have for the moving nucleon

\begin{equation}	\label{W_N}
W^{xy}_{N}= \frac{p_x p_y}{M^2} W_{2N} (p,q) +
\frac{i}{2M^2} W_{3N}[p_z q_0 -p_0 q_z]\,.
\end{equation}
Since we have
$$
q_0 W_{3A}=F_{3A}(x)\,,
$$

\begin{equation}
\frac{pq}{M} W_{3N} (p,q) = F_{3N} (x_N)\,,
\end{equation}
with $x$ as defined in Eq.~(\ref{Bj_var}) and $x_N$ is the Bjorken
variable expressed in terms of the nucleon variables , $(p^0, \bbox{p})$, in
the nucleus

\begin{equation}
x_N=\frac{Q^2}{2pq}\,,
\end{equation}
we obtain the expression for $F_{3A}(x)$ in the Bjorken limit

\begin{equation} 
\frac{F_{3 A} (x)}{A} = 4 \int d^3 r \; \int \frac{d^3 p}{(2 \pi)^3} 
\; \frac{M}{E (\bbox{p})} \; \int^{\mu}_{- \infty} \; d p^0 S_h (p^0, \bbox{p})
\; \frac{x_N}{x}
\left[\frac{p_0 q_z-p_z q_0}{Mq_z}
\right]F_{3N}(x_N)\,,
\end{equation}
where the contribution of $W_2$ appearing in Eq.~(\ref{W_N}) vanishes
after momentum integration.

Defining $\gamma$ as

\begin{equation}	\label{gamma}
\gamma=\frac{q_z}{q^0}=
\left(1+\frac{4M^2x^2}{Q^2}\right)^{1/2}\,,
\end{equation}
we get

\begin{equation} 	\label{finalF3}
\frac{F_{3 A} (x)}{A} = 4 \int d^3 r \; \int \frac{d^3 p}{(2 \pi)^3} 
\; \frac{M}{E (\bbox{p})} \; \int^{\mu}_{- \infty} \; d p^0 S_h (p^0, \bbox{p})
\left(\frac{p_0 \gamma-p_z}{(p_0-p_z) \gamma}
\right)F_{3N}(x_N)
\end{equation}

\section{Results and conclusion}

We begin by showing the results for $R_3=F_{3A}/AF_{3N}$ and comparing
them with the corresponding ratio $R_2=F_{2A}^{l}/AF_{2N}^{l}$ in Fig.~3.
For $F_{3N}$ and $F_{2N}^l$ we have taken the parametrization of \cite{MRS2}.
For the meson structure functions that contribute to $R_2$ we have
used the parametrization given in \cite{Gluck}.
We can see some similarities in the region around
$x=0.5 \sim 0.6$ where the ratio is smaller than unity and
which is mostly due to nuclear binding effects, as discussed in detail in
\cite{Akulinichev,xlt1}. Similarities in the region of $x>0.6$, where the
ratio shows a fast increase, are also apparent, and they are mostly
due to the effect of Fermi motion \cite{Bodek,xlt1}. The differences
between the neutrino and charged lepton ratios are more important
at values of $x<0.6$. These differences are mostly due to the lack
of meson renormalization effects in the neutrino structure function,
as can be seen in the figure, where we also show the results for
$R_2$ without the meson renormalization effects.

We should note that the nuclear effects are sizeable, with values of $R_3$
around 0.8 in the region of $x\simeq 0.6$ and around 0.9 for low values
of $x$. These nuclear corrections are considerably larger than those
found for the deuteron \cite{Sidorov}, as it was also the case for
$R_2$ in the charged lepton case.

In Fig.~4 we show results for $R_3$ for different values of $Q^2$.
We observe that for low values of $x$ the results for $R_3$ are rather
independent of $Q^2$, but this is not the case at large values of $x$,
where there are substantial differences. Given the fact that the most
important contribution to the GLS sum rule comes from the values
of $x<0.4$, the results of the figure indicate that nuclear effects can
reduce the GLS sum by about 10\%. Since one of the purposes of
this work is to facilitate the task of experimentalists in analyses of
the GLS sum rule and other QCD predictions for $F_{3N}$, we provide
here an easy parametrization of $R_3$ which can serve to induce the
results for $F_{3N}$ from the measured nuclear data of $F_{3A}$.
In table I we give the parameters of a fit of the ratio $R_3 (x)$ for
the $Q^2$ values of 5, 30 and 50 GeV$^2$. The form used is
$ R_3(x)=A(B-x)^{\alpha}/(1-x)^{\beta}$. For larger values of $Q^2$
results are similar to those for 50 GeV$^2$.

We also want to stress that the factor 
$(p_0 \gamma-p_z)/(p_0-p_z) \gamma$ appearing in Eq.~(\ref{finalF3})
which goes to unity in the Bjorken limit (see Eq.~(\ref{gamma})),
is not negligible for values of $Q^2 \approx 4M^2$ and produces
changes of around 10\% in $R_3$ for values of $x \approx 0.8-1$
at $Q^2=5$ GeV$^2$.

In Fig.~5 we compare our results with those of Kulagin \cite{Kulagin}
and ST \cite{Sidorov}. It is quite clear that both for the case of
Kulagin and the present results the differences with the deuteron
results of ST are significant. Our results and those of Kulagin are
qualitatively similar around $x \approx 0.4-0.5$, but they divert from each
other both at large $x$ and small $x$. The discrepancies at large
$x$ are not surprising since it was shown in \cite{quasi} that
the results for $F_{2A}$ at large $x$ were very sensitive to the
parametrization used for $F_{2N}$, which is different in \cite{Kulagin}
and the present case. These differences, however, should not be
very relevant for tests of the GLS sum rule, since the contribution to
the sum rule from the region of $x>0.6$ is very small. The discrepancies
for values of $x<0.4$ are more relevant in connection with this sum rule
since 10\% effects, typical differences at low $x$ between both
approaches, are of the same order of magnitude as the QCD
corrections to the GLS. The main reason for the discrepancies in the two
results should be seen from the different approaches followed.
In \cite{Kulagin} the spinors are normalized in a way to 
force baryon number conservation in a non relativistic formalism, in line
of the idea of the flux factor of \cite{Frankfurt} to introduce
relativistic corrections in a nonrelativistic formalism. The advantage
of the present work is that it uses a relativistic formalism from the
beginning, in which baryon number conservation is automatically
fulfilled. The differences of 10\% in the region of small $x$ between
the two aproaches is about the same one as found between our approach
and those using the flux factor in studies of $F_{2A}^{l}$
\cite{xlt1,Ciofi}.

In summary, we present here results for $R_3$ based on a relativistic
approach and the use of an accurate nuclear spectral function which
reproduces nuclear bindings, momentum distributions, etc. Within this
formalism we were able to reproduce the EMC effect for different nuclei
and the experimental results of $F_{2A}$ in the region of $x>1$. We have used
here the same formalism to evaluate $F_{3A}$. There are some differences
with respect to $F_{2A}$ in the nuclear effects, mostly due to the
absence of meson renormalization corrections in $F_{3A}$. The results
obtained are parametrized in an easy way to facilitate future
analyses of experimentalists in order to deduce the elementary $F_{3N}$
structure function from the nuclear measurements and be able to test
QCD corrections and obtain reliable values of $\alpha_s$.

\acknowledgements

We would like to acknowledge W.~Seligman for providing us with a copy
of his Ph.~D. Thesis and interesting comments. One of us, S.K.S. wishes
to acknowledge financial support from Ministerio de Educaci\'on y Cultura
in his sabbatical stay in Valencia. E.M. wishes to acknowledge financial
support from the Ministerio de Educaci\'on y Cultura. This
work is partially supported by DGICYT, contract no.
PB96-0753.

\begin{table}
\caption{Fits of the ratio $R_3(x,Q^2)=F_{3A}(x,Q^2)/AF_{3N}(x,Q^2)$
for different values of $Q^2$. The form of the fit is 
$R_3(x)=A(B-x)^{\alpha}/(1-x)^{\beta}$
\label{table1}}
\begin{tabular}{ccccc}
$Q^2$ (GeV$^2$) & $A$ & $B$ & $\alpha$ & $\beta$ \\
\tableline
5  & 0.691 & 1.29 & 1.21 & 0.688\\
30 & 0.647 & 1.06 & 6.88 & 6.19 \\
50 & 0.599 & 1.07 & 6.43 & 5.64 \\
\end{tabular}
\end{table}

\begin{figure}
\caption{Diagram for the process of deep inelastic neutrino-nucleon
scattering.
\label{fig1}}
\end{figure}

\begin{figure}
\caption{Self-energy diagram of the neutrino in the nuclear medium
associated with the process of deep inelastic neutrino-nucleon
scattering.
\label{fig2}}
\end{figure}

\begin{figure}
\caption{ Solid line: results for the ratio $F_{3A} (x)/A F_{3N} (x)$;
dashed line: results for the ratio $F_{2A}^l (x)/A F_{2N}^l (x)$
including the contribution of the nucleons and the mesons;
dotted-dashed line: results for the ratio $F_{2A}^l (x)/A F_{2N}^l (x)$
including only the contribution of the nucleons. All curves are evaluated
at $Q^2 = 50 $ GeV$^2$.
\label{fig3}}
\end{figure}

\begin{figure}
\caption{Results for the ratio $F_{3A} (x)/A F_{3N} (x)$ at different values
of $Q^2$. Dotted-dashed line: ratio at $Q^2= 50$ GeV$^2$; long
dashed line: ratio at $Q^2= 30$ GeV$^2$; solid line: ratio at
$Q^2= 5$ GeV$^2$; short dashed line: ratio at $Q^2= 5$ GeV$^2$ but setting
$\gamma$ equal to 1 in Eq.(\protect\ref{finalF3}).
\label{fig4}}
\end{figure}

\begin{figure}
\caption{Results for the ratio $F_{3A} (x)/A F_{3N} (x)$ at $Q^2=5$ GeV$^2$
by different authors. Solid line: this work; dashed line: Kulagin 
\protect\cite{Kulagin}; dotted-dashed line: Sidorov and Tokarev
\protect\cite{Sidorov} (for deuteron).
\label{fig5}}
\end{figure}

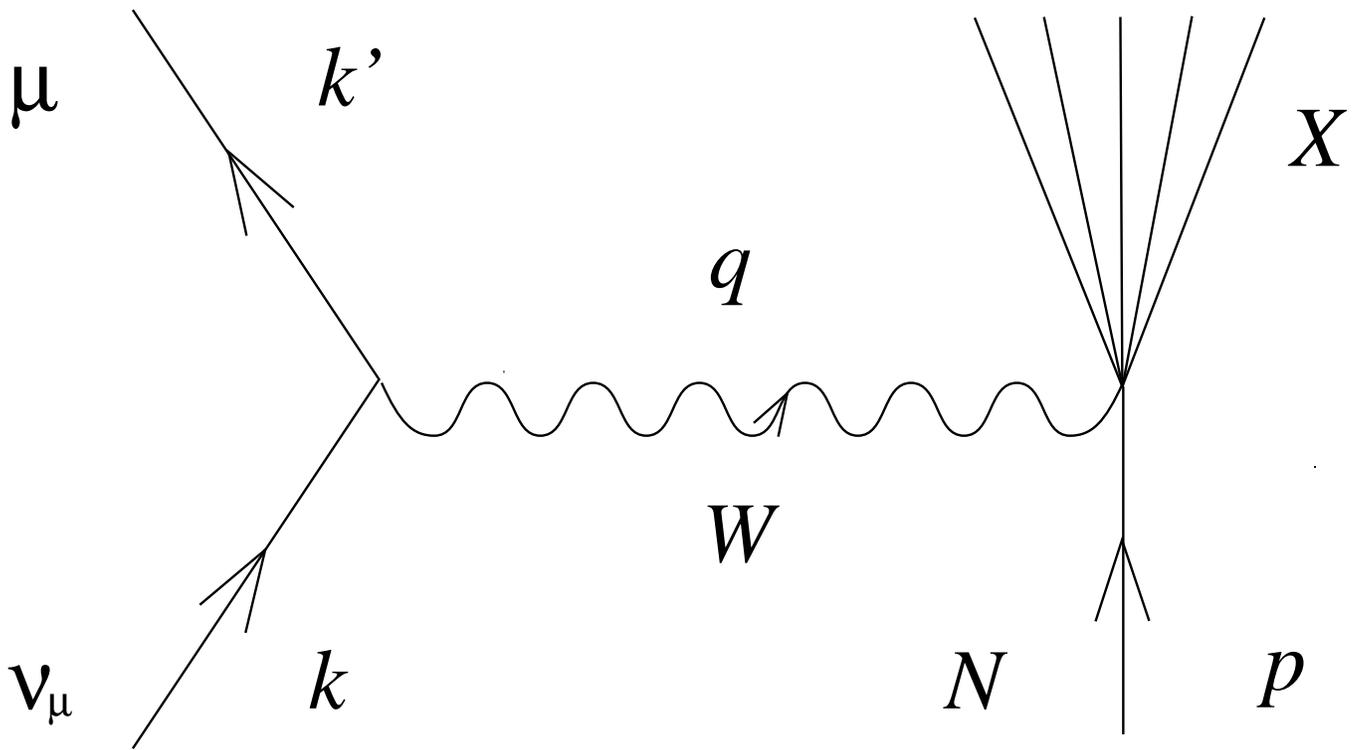
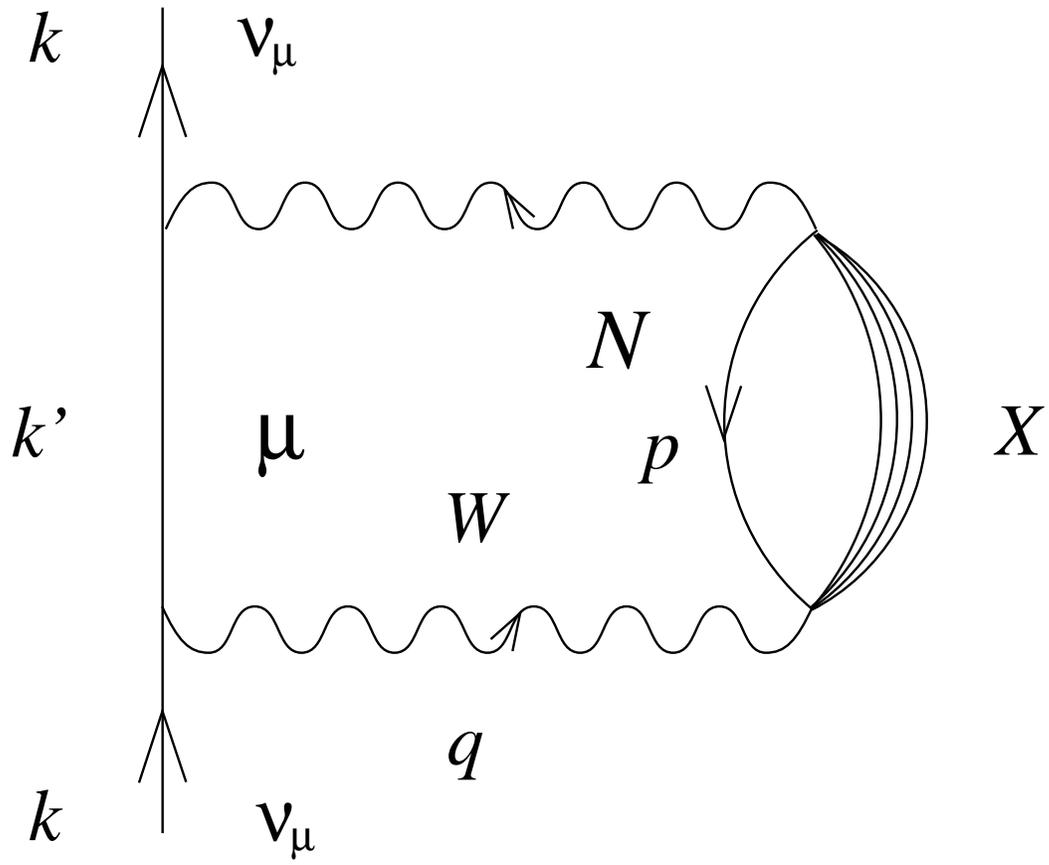
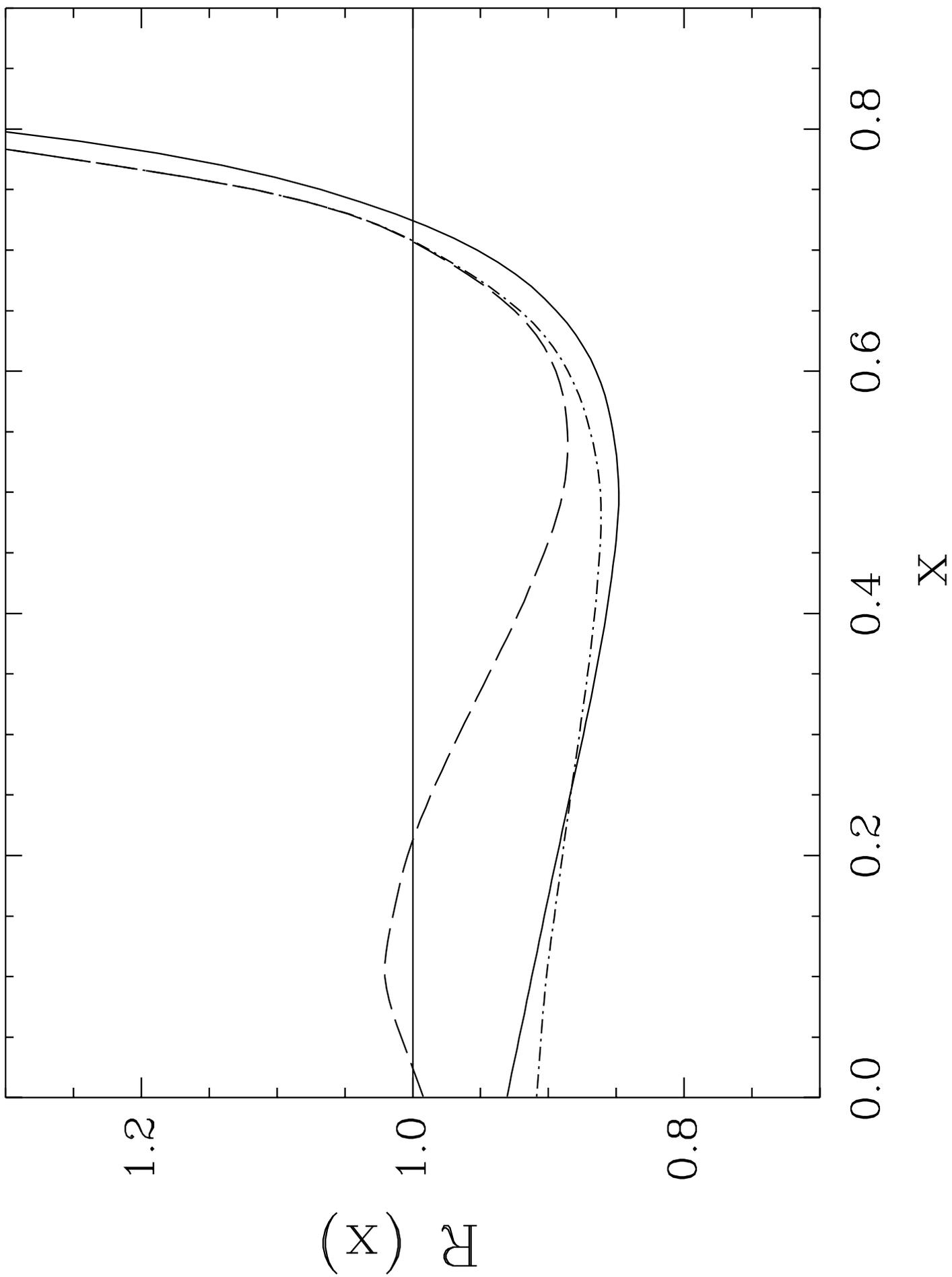
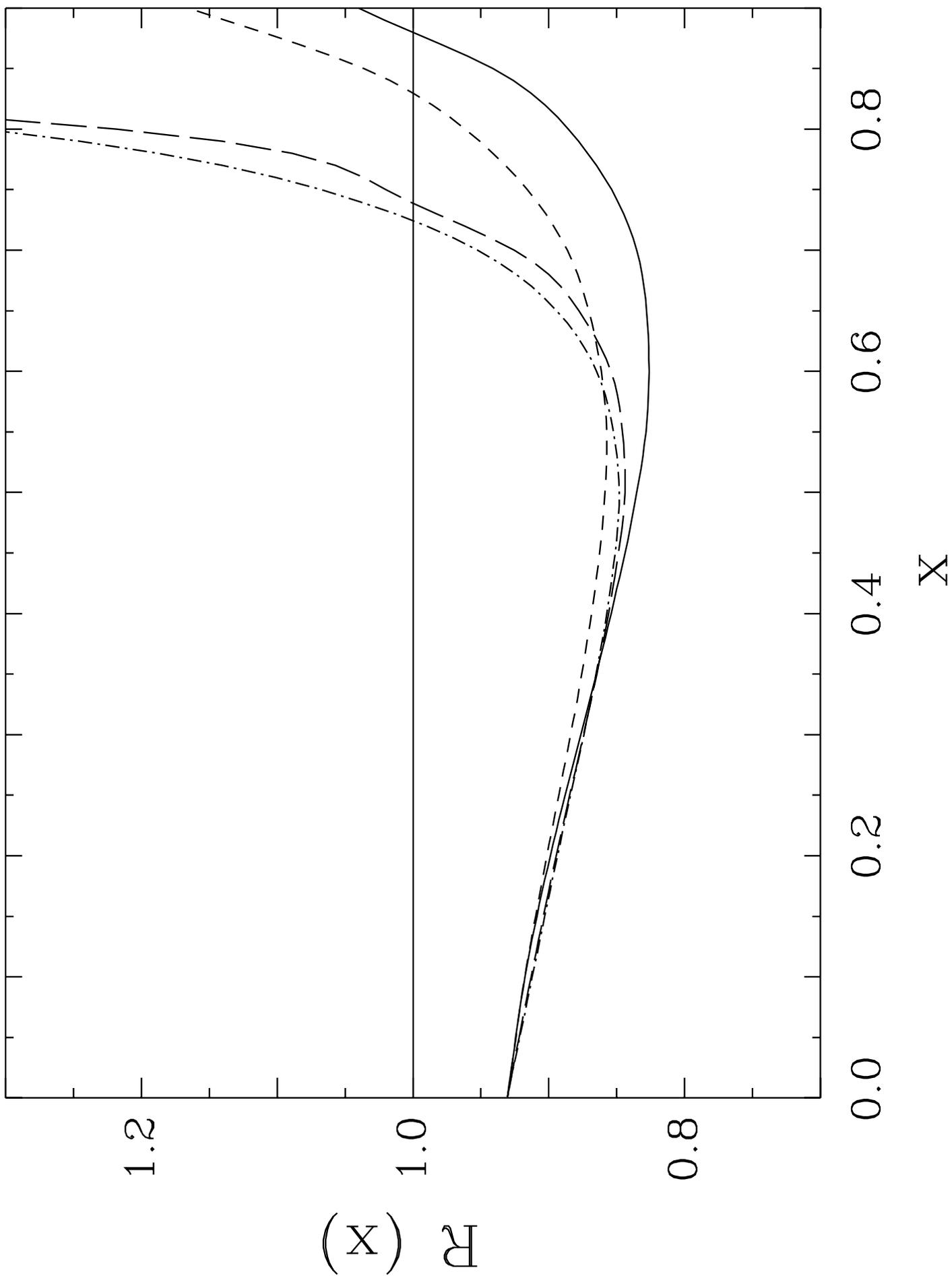
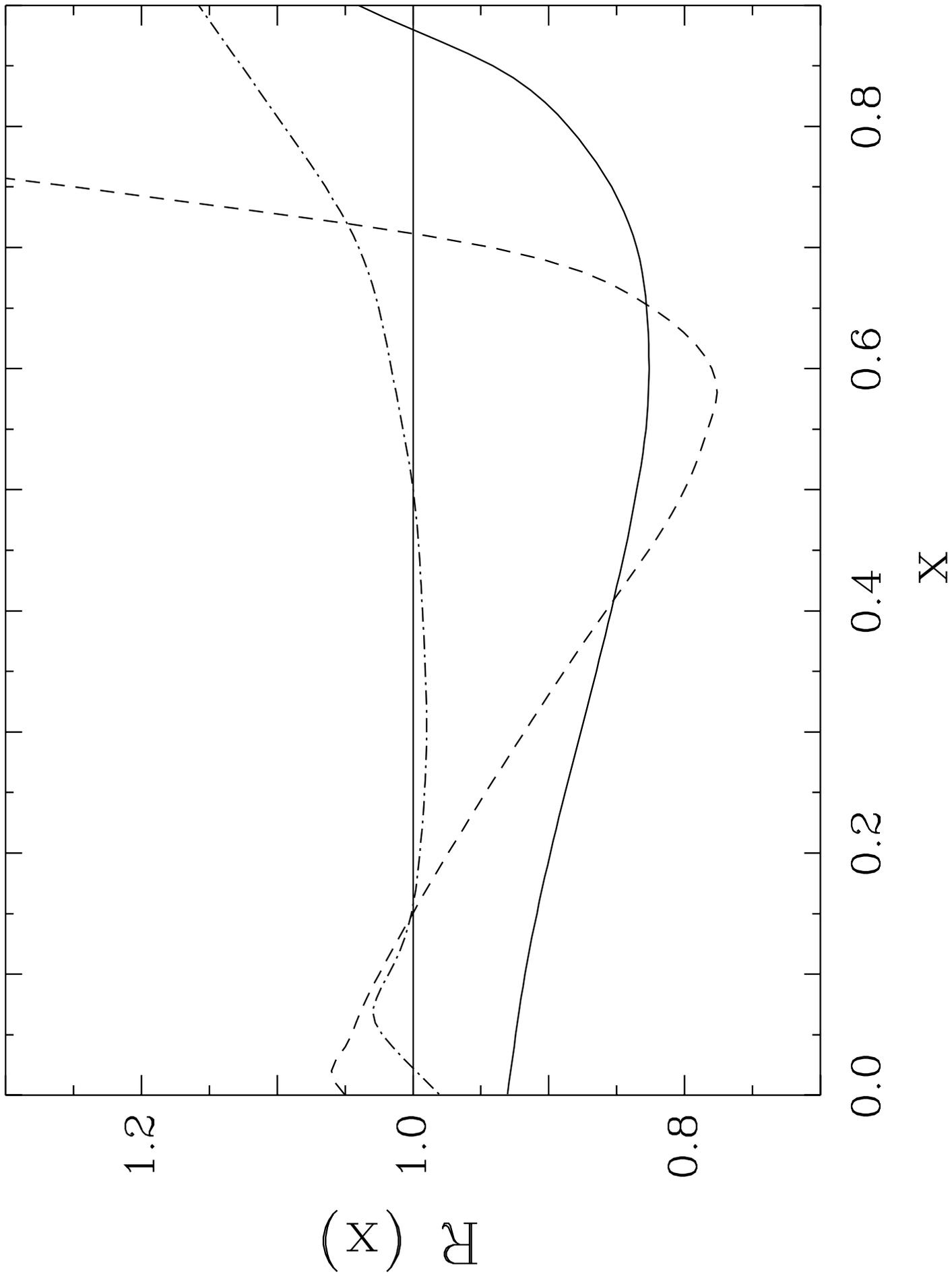
\end{document}